\documentstyle[twocolumn,prl,aps]{revtex}
\begin{document}
\draft
\title{Self-Organized Criticality:  Self-Organized Complexity?  
\\ The Disorder and ``Simple Complexity'' of Power Law Distributions}
\author{J.S.~Shiner\thanks{Mailing address: Bergacher 3, CH-3325 
Hettiswil, Switzerland. Phone: + 41 34 411 02 43.  Electronic 
address: shiner@alumni.duke.edu.}}
\address{Physiologisches Institute, Universit\"at Bern, Switzerland}
\date{\today}
\maketitle
\begin{abstract}
The disorder and a simple convex measure of complexity are studied for 
rank ordered power law distributions, indicative of criticality, in 
the case where the total number of ranks is large.  It is found that a 
power law distribution may produce a high level of complexity only for 
a restricted range of system size (as measured by the total number of 
ranks), with the range depending on the exponent of the distribution.  
Similar results are found for disorder.  Self-organized criticality 
thus does not guarantee a high level of complexity, and when 
complexity does arise, it is self-organized itself only if 
self-organized criticality is reached at an appropriate system size.
\end{abstract}


Self-organized criticality \cite{Baketal1987} has been maintained to 
be ``so far the only known general mechanism to generate complexity'' 
\cite{Bak1996}.  Actually, self-organized criticality says nothing 
about complexity itself directly.  Rather, it can be taken as defined 
by two properties.  The first is that of the critical state, indicated 
by a power law probability distribution.  The second, 
self-organization, is that the critical state is reached 
spontaneously, without the influence of any external agent, to set 
parameter values for example.  The question asked here is whether a 
power law distribution necessarily implies complexity.

Power law distributions usually occur in one of two forms.  In the 
first the probability $p(m)$ of an event of magnitude $m$ is expressed 
as a power of the magnitude itself: $p(m) \propto 1/{m}^{\gamma }$, 
where $\gamma$ is a nonnegative constant.  Examples include the 
Gutenberg-Richter law relating the frequency of earthquakes to the 
energy released \cite{GutenbergRichter1949,JohnstonNava1985}, 
fluctuations in the stock and commodities markets 
\cite{Mandelbrot1963,Mandelbrot1964}, and the number of extinctions 
throughout biological evolution \cite{RaupSepkoski1984,Raup1986}.  It 
should be noted that some examples are not uncontroversial; for 
example, there are alternative explanations for the frequencies of 
biological extinctions \cite{Newman1996}.

In the second common form of power law distributions the distributions 
are rank ($r$) ordered.  The most frequent or probable event or state 
has rank $1$; the second most probable, rank $2$; {\it etc.} The 
distribution is then written $p(r) \propto 1/{r}^{\gamma }$.  The 
first well known example was Zipf's law for the distribution of word 
frequencies in the English language \cite{Zipf1949}; others are the 
distribution of city populations and various aspects of the ``web''  
\cite{glassman,hubermanandadamic}.  Since the rank ordered form of power law 
distributions is the marginally simpler one, it will be used in the 
following.

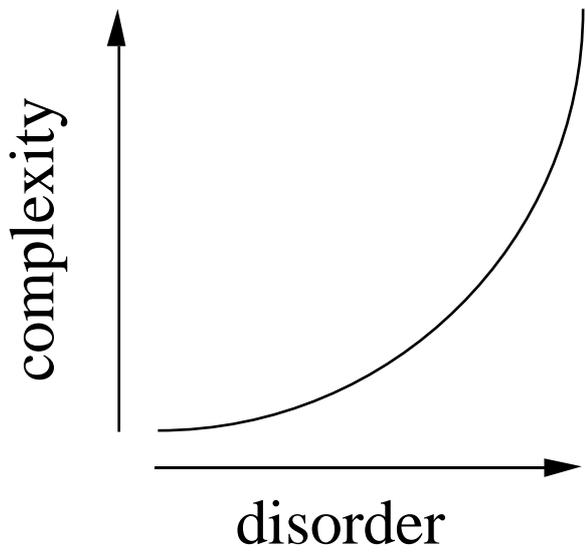
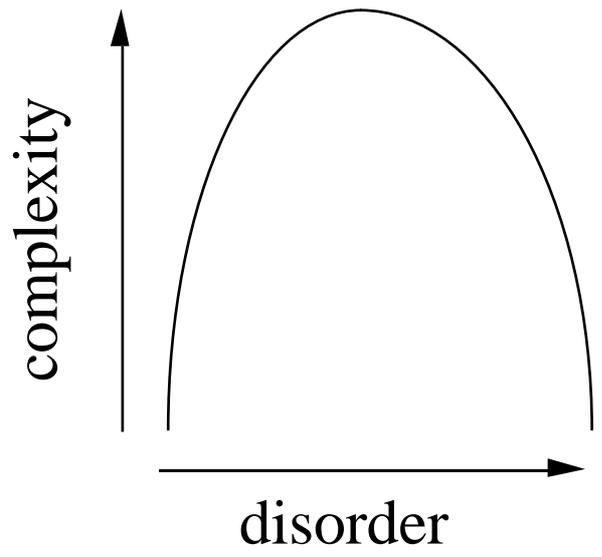
\begin{figure}
	\caption{Classification of complexity measures based on dependence 
on disorder.  Left: complexity increases monotonically with 
disorder; right: complexity displays a convex dependence on 
disorder.}
	\label{fig1}
\end{figure}

There is a plethora of proposals for complexity measures in the 
literature, each trying to capture one or more aspects of what we mean 
when we say that something is complex.  Most current measures may be 
assigned to one of two categories 
\cite{Wackerbaueretal1994,Atmanspacher1994,Atmanspacheretal1997}, 
Fig.~\ref{fig1}, \cite{caveat}.  In the first, complexity increases 
monotonically with disorder.  In the second, complexity shows a convex 
dependence on disorder, vanishing at minimum and maximum disorder, and 
obtaining a maximum at an intermediate level of disorder.

It is important to recognize that entropy is not necessarily a good 
measure for disorder, {\em e.g.} when the number of states available 
to a system can change.  For this reason Landsberg 
\cite{Landsberg1984a,Landsberg1984b,Landsberg1994} proposed that 
disorder $\Delta $ be taken as
\begin{equation}
\Delta \equiv   S/{S}_{\max},\ S=-\sum\limits_{  i=1}^{  N} 
{  p}_{  i}  \ \ln {p}_{i},
	\label{eq1}
\end{equation}
where ${p}_{i}$ is the probability of state $i$ of $N$ possible 
states, and $S$ is therefore the information entropy.  $S_{max}$ is the 
appropriate maximum entropy.  This measure for disorder has been 
applied to problems ranging from cosmology \cite{Landsberg1994} to 
biology \cite{Shiner1997}, and  can be generalized easily
\cite{Shineretal1999a} to measures based on higher order R\'{e}nyi 
entropies \cite{Renyi1970}, which are related to multifractals 
\cite{Grassberger1983,HentschelProcaccia1983}.

Based on the Landsberg ``disorder'' \cite{quotes} we have proposed a 
``simple measure of complexity'' 
\cite{Shineretal1999a,Shineretal1999b} whose behavior encompasses both 
the monotonic and convex categories:
\begin{equation}
	{\Gamma }_{\alpha \beta }\equiv {\Delta }^{\alpha }{\left({  
1-\Delta }\right)}^{\beta }  ,\ \alpha   \ \mathrm{and} \ \beta   
\ \mathrm{constants}.
	\label{eq3}
\end{equation}
When $\alpha >0$ and $\beta $ vanishes, ${\Gamma }_{\alpha \beta }$ is 
a monotonic complexity measure; when $\alpha ,\beta >0$, ${\Gamma 
}_{\alpha \beta }$ is a convex measure.  Since the results here 
will not depend on the particular dependence of complexity on 
disorder, but rather only on whether the dependence is monotonic or 
convex, in the following only the simplest cases will be considered:
\begin{equation}
\matrix{\alpha =1,\ \beta =0:\ {\Gamma }_{10}=\Delta ;\rm \ 
monotonic;\cr \cr \alpha =1,\ \beta =1:\ {\Gamma 
}_{11}=\Delta \left({1-\Delta }\right) \equiv \Gamma;\rm \ convex.\cr}
	\label{eq4}
\end{equation}
For convenience these will be referred to simply as ``disorder'' and 
``complexity'', respectively.

It is important to understand that $\Delta$ is not equivalent to the 
entropy $S$ and that, therefore, $\Gamma$ is not simply a function of 
entropy.  To say that $\Gamma$ is a function of $S$ alone is to 
misunderstand the reason for introducing $\Delta$ as an alternative to 
entropy as a measure of disorder in the first place 
\cite{Landsberg1984a,Landsberg1984b,Landsberg1994}.  As $S$ varies, 
$S_{max}$ will in general also vary, but not proportionally to $S$, so 
that $\Delta$ will in general change in a manner distinct from the 
variation in $S$.  $\Delta$ may even decrease as $S$ increases.  Both 
$S$ and $S_{max}$ will generally change (nonproportionally) in several 
scenarios, including changes in system size, changes in the level of 
fine graining and changes in boundary conditions.  Moreover, there may 
be several choices for $S$ and $S_{max}$ for a given system.  Which 
ones are appropriate will depend on the questions being addressed.  In 
the simplest cases $S_{max}$ can be taken to be the entropy of the 
equiprobable (all $p_{i}=1/N$) distribution, $\ln N$.  For 
nonequilibrium systems, it {\em may} be more appropriate to take 
$S_{max}$ to be the entropy of the corresponding equilibrium system, 
{\em i.e.}~the equilibrium system with the same total energy, number 
of particles, {\em etc.} \cite{noneqgas}.  Even for a simple 
one-dimensional Ising spin system, several entropies and therefore 
several ``disorders'' can be introduced 
\cite{Shineretal1999b,replytocomments}.  Furthermore, multiple choices 
for $S$ and $S_{max}$ imply multiple $\Delta$'s and multiple 
$\Gamma$'s.  When one considers that $\Gamma$ calculated from one 
$\Delta$ can be investigated as a function of another ``disorder'', one 
recognizes that eqs.~(\ref{eq3},\ref{eq4}) allow for a multiplicity 
of disorder-complexity relations.  In short, eq.~(\ref{eq3}) is the 
definition of ${\Gamma}_{\alpha \beta}$ and does not uniquely 
determine its dependence on $\Delta$.  For a discussion of these 
points and eq.~(\ref{eq3}) as a measure of complexity see 
\cite{CFScomment,BPcomment,replytocomments}.

For the distributions studied here only one $S_{max}$ will be 
investigated, that of the equiprobable distribution, and ${\Gamma}$ is 
determined by $\Delta$.  This is appropriate for distinguishing 
between complexity measures which increase monotonically with disorder 
and those which show a convex dependence on disorder, one purpose of 
this study.

The entropy, ``disorder'' and ``complexity'' of the rank ordered power law 
distribution, $p(r) \propto   1/{r}^{\gamma }$, with maximum rank $R$ 
are now
$$p(r)=1/\left({\Xi { r}^{\gamma }}\right),\ \sum\limits_{ 
r=1}^{ R}  p(r)=1\ \Rightarrow  \ \Xi  =\sum\limits_{ 
r=1}^{ R}  1/\left({{r}^{\gamma }}\right),$$
\begin{equation}
S=-\sum\limits_{ r=1}^{ R}  p(r) \ln p(r)=\ln\Xi  +\left({\gamma 
/\Xi }\right)\sum\limits_{ r=1}^{ R} \left({ \ln r}\right)/{ 
r}^{\gamma } ,
	\label{eq5}
\end{equation}
$${S}_{\max}=\ln R,\ \Delta  =S/{S}_{\max},\ \Gamma  =\Delta 
\left({ 1-\Delta }\right) .$$
$R$ may have different interpretations.  The most straightforward  
is that of $R$ as the maximum observed rank.  $R$ may also be taken 
to be the size of the system, in the sense that a larger system has 
more states, {\it i.e.}~more ranks.  Finally, $R$ may be interpreted 
as a measure of the extent of fine graining.  Examples of power law 
distributions are often presented as histograms, {\it e.g.}~the 
number of biological extinctions $<10\%$, between $10$ and $20\%$, 
\ldots \cite{Bak1996}.  If the distribution is more refined, say $<1\%$, 
between $1$ and $2\%$, \ldots , then $R$ obviously increases.  In the 
limit $R\rightarrow \infty $ we have the fine grained or 
thermodynamic limit.

Since the latter limit is of particular importance, I will now 
consider the case of large $R$.  Replacing sums by integrals, 
we have \cite{Riemann}
$$\matrix{p={1 \over \Xi { r}^{\gamma }},\ \int_{1}^{R}p\ \mit d 
\,r=1\ \cr \Rightarrow  \ \Xi  =\left\{{\matrix{\ln R,&\gamma 
 =1,\cr \left[{{R}^{\left({1-\gamma 
}\right)}-1}\right]/\left({1-\gamma }\right),&\gamma \ne  
1,\cr}}\right.\cr}$$
\begin{equation}
\matrix{S=-\int_{1}^{R}p\ \ln p\ \mit d \,r=\cr \cr 
\left\{{\matrix{\ln \left({\ln R}\right)+{\ln R \over 2},&\gamma 
 =1,\cr \ln \left[{{{R}^{\left({1-\gamma }\right)}-1 \over 
\left({1-\gamma }\right)}}\right]+{\gamma { R}^{\left({ 1-\gamma 
}\right)} \ln R \over {R}^{\left({1-\gamma }\right)}-1}-{\gamma 
\over \left({ 1-\gamma }\right)},&\gamma \ne  
1,\cr}}\right.\cr}
	\label{eq6}
\end{equation}
$${S}_{\max}=\ln\left({R-1}\right).$$

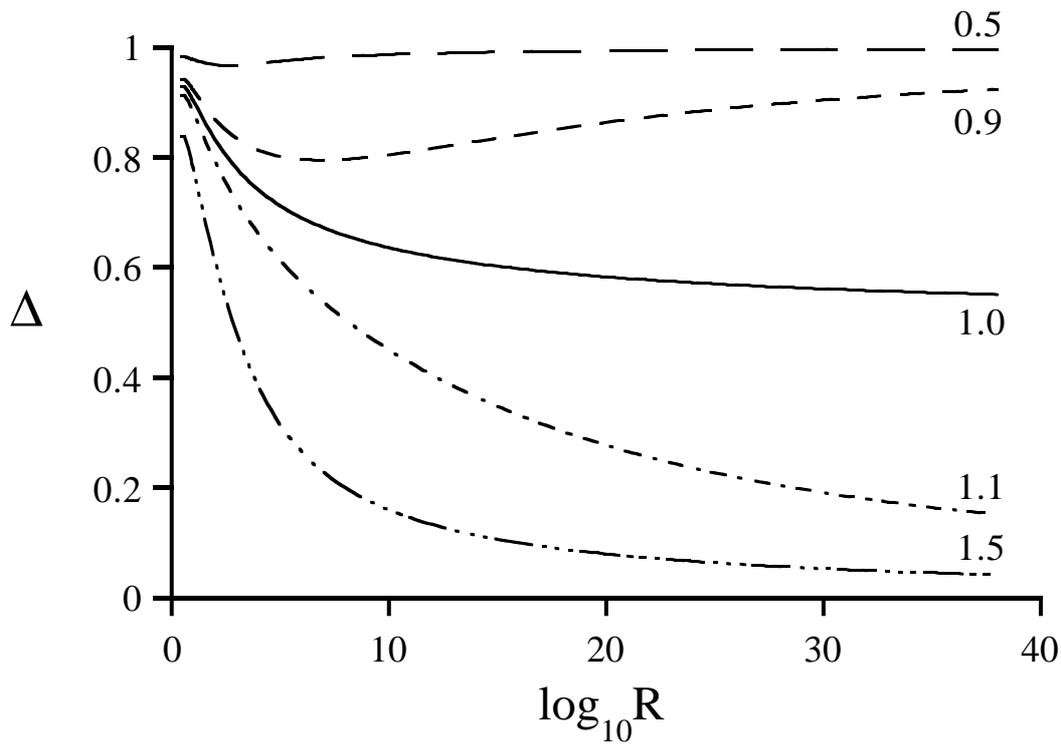
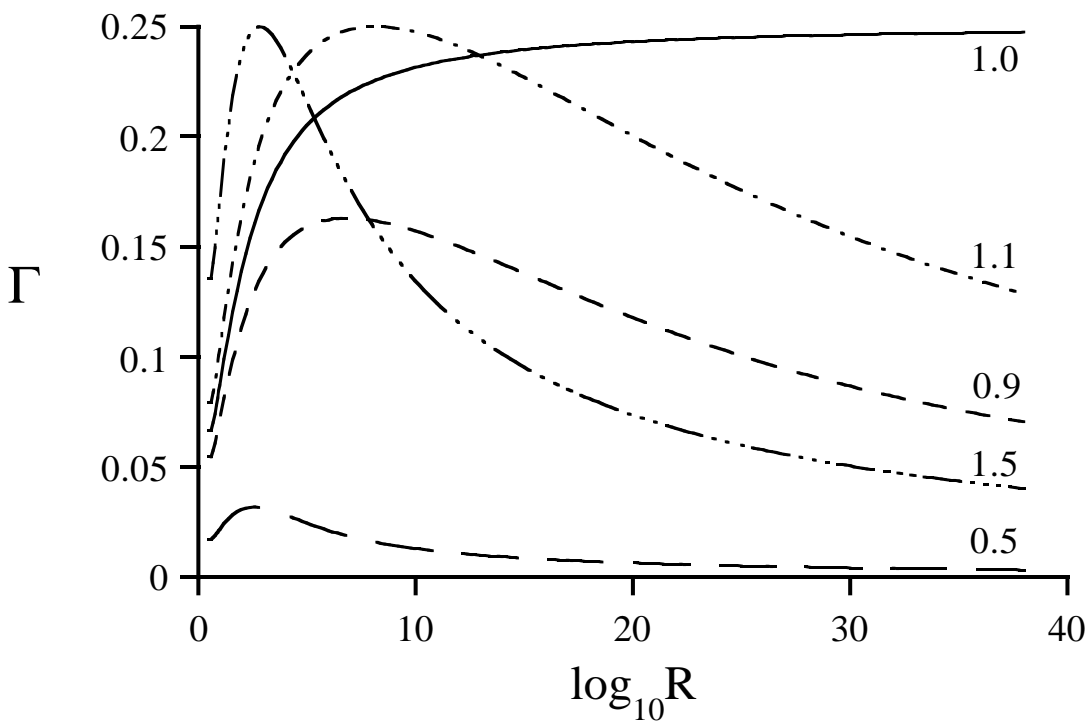
\begin{figure}
	\caption{The dependence of ``disorder'' and ``complexity'' on the maximum 
rank $R$.  The curves are labeled with the values of the exponent 
$\gamma $.}
	\label{fig2}
\end{figure}

The dependence of ``disorder'' and ``complexity'' on $R$ is shown in  
Fig.~\ref{fig2}.  The first point to note is
\begin{equation}
\lim\limits_{R \rightarrow \infty }^{}\ \Delta   
=\left\{{\matrix{1,\cr
1/2,\cr
0,\cr}\matrix{\gamma   <1\cr
\gamma   =1\cr
\gamma   >1\cr}}\right\},\ \lim\limits_{R\rightarrow   \infty 
}^{}\ \Gamma   =\left\{{\matrix{0,\cr
1/4,\cr
0,\cr}\matrix{\gamma   <1\cr
\gamma   =1\cr
\gamma   >1\cr}}\right\}.
	\label{eq7}
\end{equation}
In the fine grained limit, the system is either completely disordered 
($\gamma <1$) or completely ordered ($\gamma >1$), with the one 
exception of $\gamma =1$, where the system is half maximally 
disordered (half maximally ordered).  In the thermodynamic limit 
``complexity'' vanishes except for the special case $\gamma =1$, where 
``complexity'' has its maximum possible value (with the choice of 
$\alpha = \beta =1$).  In this limit self-organized criticality 
does not imply ``complexity'' in general: ``complexity'' itself vanishes 
except for $\gamma =1$, and ``disorder'' vanishes except for $\gamma \le 
1$.

For $R$ large, but $R<\infty $, significant points to be noted are:
\begin{itemize}
\item[] for $\gamma < 1$, $\Delta$ goes through a relative minimum and 
$\Gamma$ through a relative maximum as $R$ increases;
\item[] for $\gamma > 1$,   $\Delta$ decreases monotonically with $R$, and 
$\Gamma$ first increases to its absolute maximum to then vanish in the 
limit $R\rightarrow \infty $.
\end{itemize}
Values of $\gamma $ other than $1$ may yield nonvanishing values of 
``complexity'', but only for some range of intermediate values of $R$.  
It should be noted that these results are not limited to the 
particular forms for disorder and complexity used here.  Qualitatively 
the complexity results apply for all convex complexity measures, since 
they all vanish for minimum and maximum disorder.  Similarly, the 
disorder results are valid qualitatively for all monotonic complexity measures.

Even if the power law probability distribution of self-organized 
criticality does lead to a high level of ``complexity'', this does not 
imply that the ``complexity'' itself is self-organized.  For 
self-organized complexity, the system must not only evolve to a power 
law probability distribution spontaneously, it must also do so so that 
$\gamma $ is exactly $1$ in the limit $R\rightarrow   \infty $.  Or, 
if $R$ is large but finite, the system must spontaneously evolve to a 
size for which the complexity is large for the operative value of 
$\gamma $.  Thus self-organized complexity requires much more 
stringent conditions than self-organized criticality.  Whether the 
many systems which are known to be examples of self-organized 
criticality also show self-organized complexity remains to be seen.

Two additional points --- (1) Mandelbrot's \cite{Mandelbrot1961} 
finding, that a monkey pounding on a typewriter at random produces a 
rank ordered power law distribution for the ``words'' typed, is 
consistent with the result of maximum ``disorder'' and vanishing 
``complexity'' in the fine grained limit for $\gamma <1$.  (2) Bak 
\cite{Bak1996} alludes to the ``blind watchmaker'' argument of 
Dawkins \cite{Dawkins1988}.  This is 
essentially an anti-vitalist argument, namely that the order and 
complexity produced by evolution apparent to us do not require the 
intervention of an external agent, {\it i.e.}~``divine'' intervention.  
In the case of self-organized criticality the argument is that nature 
can produce criticality and therefore complexity without the 
intervention of an external agent.  However, 
self-organized criticality produces a high level of ``complexity'' in 
the thermodynamic limit only for $\gamma =1$, or for $R$ large but 
finite, only for certain system sizes.  For the ``blind watchmaker'' 
argument to apply to ``complexity'', one must show not only that 
criticality is reached in a self-organized fashion but also that it is 
reached at an appropriate system size and $\gamma $.  Otherwise, if 
the system size necessary for a high level of ``complexity'' is not 
reached spontaneously but must be set by an outside agent, the 
``watchmaker'' must have sufficient visual acuity to do so.

\begin{references}
\bibitem{caveat}This classification scheme may not encompass all 
	proposed complexity measures, however; see \cite{replytocomments}.
\bibitem{quotes}When referring to specific measures of disorder, order 
  or complexity, we enclose these words in quotation marks to 
  distinguish the measures from the general concepts.
\bibitem{Riemann}Replacing the discrete formulation of eq.~(\ref{eq5}) 
with the continuous approximation of eq.~(\ref{eq6}) can lead to 
spurious results in certain cases.  In particular, in the fine grained 
limit, $R\rightarrow \infty $, the continuous entropy of 
eq.~(\ref{eq6}) can go negative for $\gamma$ sufficiently large.  In 
this limit, the discrete case can be represented exactly in terms of 
the Riemann zeta function, 
$\zeta \left({ x}\right)\equiv 
\lim\limits_{R\rightarrow \infty }^{}{\sum}_{r=1}^{R} 1/{r}^{x}$: 
$S=\ln \zeta 
\left({\gamma }\right) -\gamma \left[{\partial \,\zeta \left({\gamma 
}\right)/\partial \,\gamma }\right]/\zeta \left({\gamma 
}\right)\rm .$  
This equation yields the same 
results for $\Delta$ as does the continuous approximation of 
eq.~(\ref{eq6}).
\end{references}
\end{document}